\documentclass[12pt]{article}
\usepackage{geometry}                % See geometry.pdf to learn the layout options. There are lots.
\usepackage{graphicx}
\usepackage{amssymb}
\usepackage{epstopdf}
\usepackage{color}

\usepackage{amsmath}
\usepackage{graphics,graphicx,epsfig}
\usepackage{amssymb}
\usepackage{sidecap}
\sidecaptionvpos{figure}{c}
\usepackage{appendix}

\def\nbar{N_{CFT}}

\begin{document}

\begin{titlepage}
\vfill
\begin{flushright}
ACFI-T14-17\\
%\today
\end{flushright}

%\vfill
\vskip 3.0cm
\begin{center}
\baselineskip=16pt
{\Large\bf Chemical Potential in the First Law for Holographic Entanglement Entropy}

\vskip 10.mm

{\bf  David Kastor${}^{a}$, Sourya Ray${}^{b}$, Jennie Traschen${}^{a}$} 

\vskip 0.4cm
${}^a$ Amherst Center for Fundamental Interactions, Department of Physics\\ University of Massachusetts, Amherst, MA 01003\\

\vskip 0.1in ${}^b$ Instituto de Ciencias F\'{\i}sicas y Matem\'{a}ticas, Universidad Austral de
Chile, Valdivia, Chile\\
\vskip 0.1 in Email: \texttt{kastor@physics.umass.edu, ray@uach.cl, traschen@physics.umass.edu}
\vspace{6pt}
\end{center}
\vskip 0.2in
\par
\begin{center}
{\bf Abstract}
 \end{center}
\begin{quote}
Entanglement entropy in conformal field theories is known to satisfy a first law.  For spherical entangling surfaces, this has been shown to follow via the AdS/CFT correspondence and the holographic prescription for entanglement entropy from the bulk first law for Killing horizons. The bulk first law
can be extended to include variations in the cosmological constant $\Lambda$, which we established in earlier work.
Here we show that this implies an extension of the boundary first law to include varying the number of degrees of freedom of the boundary CFT.  The thermodynamic potential conjugate to $\Lambda$ in the bulk is called the thermodynamic volume and has a simple geometric formula.  In the boundary first law it plays the role of a chemical potential.  For the bulk minimal surface $\Sigma$ corresponding to a boundary sphere, the thermodynamic volume is found to be proportional to the area of $\Sigma$, in agreement with the variation of the known result for entanglement entropy of spheres.  The dependence of the CFT chemical potential on the entanglement entropy and number of degrees of freedom is similar to how the thermodynamic chemical potential of an ideal gas
depends  on entropy and particle number.

   \vfill
\vskip 2.mm
\end{quote}
\hfill
\end{titlepage}

%\maketitle
%\section{}
%\subsection{}

\section{Introduction}

Recently there has been great interest in how information regarding entanglement of the quantum state of a conformal field theory  is encoded, through the AdS/CFT correspondence, in the geometry of a bulk, asymptotically anti-deSitter spacetime.  According to the Ryu-Takayanagi proposal \cite{Ryu:2006bv}, the entanglement entropy $S_{E}$ associated with the division of the boundary into complementary regions $A$ and $B$ is given by the Bekenstein-Hawking entropy formula
\begin{equation}\label{entropy}
S_E={A_\Sigma\over 4 G}
\end{equation}
applied to a bulk minimal surface $\Sigma$ whose boundary at spatial infinity matches onto the boundary between $A$ and $B$ and whose area is $A_\Sigma$.  Although entanglement is not a thermal property, entanglement entropy nevertheless obeys a first law relation with respect to variations in the quantum state of the CFT  \cite{Blanco:2013joa,Wong:2013gua}.  Moreover, in the important case that the entangling surface on the boundary is a sphere, the first law for entanglement entropy has been shown to follow from a bulk gravitational first law associated with the minimal surface $\Sigma$ \cite{Faulkner:2013ica}.  This works because the surface $\Sigma$, in this case, is the bifurcation surface of a Killing horizon.   The proof of the first law for stationary black holes \cite{SudarskyWald:1992} then applies in this non-black hole setting as well.

We will extend this map between bulk and boundary first laws to show that  another feature of CFT entanglement,  namely the dependence of entanglement entropy on the number of CFT degrees of freedom, is also encoded in the bulk geometry (at least in the case of spherical entangling surfaces to which our construction applies).   In the AdS/CFT correspondence,  the number of degrees of freedom in the boundary theory determines the bulk cosmological constant $\Lambda$.  Hence, varying the number of boundary degrees of freedom corresponds to varying $\Lambda$ in the bulk.   In earlier work \cite{KastorEtal:2009} we have shown that the derivation \cite{SudarskyWald:1992} of the first law for stationary black holes can be extended to include variations in $\Lambda$.  In this 
work we will apply the
extended first law  to the bulk minmal surface/Killing horizon $\Sigma$ associated with a spherical entangling region on the boundary, resulting in an extension of the first law for $S_E$ to include variations in the number of boundary degrees of freedom $N_{CFT}$.  The new conjugate thermodynamic potential appearing in the extended first law plays the role of a chemical potential, and we will see that its dependence on $S_E$ and $N_{CFT}$ is similar to that of the thermodynamic chemical potential for an ideal gas on the entropy and particle number.
The extended first law for black holes can be stated in terms of varying the pressure $P=-\Lambda/8\pi G$ and the conjugate thermodynamic volume $V$, 
\begin{equation}
dM = TdS_{BH} +VdP
\end{equation}
where $T=\kappa/2\pi$ is the Hawking temperature, $S_{BH}=A_H/4G$ is the Bekenstein-Hawking entropy, $\kappa$ and $A_H$ are the surface gravity and black hole horizon area,  and we have suppressed terms that would arise from varying the angular momentum and charge.  The thermodynamic volume $V$ is given by a simple geometric formula involving the anti-symmetric Killing potential $\omega^{ab}$ for the Killing vector that generates the horizon.  In the simplest example of the $D$-dimensional Schwarzschild-AdS spacetime, one finds that 
\begin{equation}
V ={ \Omega_{D-2}r_h^{D-1}\over D-1}
\end{equation}
where $r_h$ is the horizon area and $\Omega_{D-2}$ is the area of a unit $D-2$ sphere.
This exactly matches the naive volume inside the horizon computed as an integral from $r=0$ to $r_h$ with the full $D$-dimensional volume element.  The properties of black holes with $\Lambda$ considered as a thermodynamic variable have received considerable attention (see the recent review \cite{Dolan:2014jva} and references therein).  In particular, the thermodynamic volume $V$ was computed for a wide range of AdS black holes in \cite{Cvetic:2010jb}.  It was found that while the thermodynamic volume of RN-AdS black holes continues to match the geometric volume inside the horizon, as described above, this fails for rotating black holes and also for more general charged black holes in gauged supergravities.  All known cases, however, satisfy an intriguing `reverse isoperimetric inequality' that bounds the thermodynamic volume from below in terms of the horizon area \cite{Cvetic:2010jb}.  Also notably, reference \cite{Kubiznak:2012wp} showed that the phase transition between large and small RN-AdS black holes has the Van der Waals form when analyzed in terms of isotherms in the $PV$-plane.  A number of works \cite{KastorEtal:2009,Johnson:2014yja,Dolan:2014cja} have also sought to apply the extended first law  for AdS black hole, in the AdS/CFT context, to properties of CFT's at finite temperature. The  role of the thermodynamic volume for deSitter black holes was 
studied in \cite{Dolan:2013ft} .

The structure of this paper will be as follows.   In section (\ref{holosection}) we briefly review the holographic prescription for entanglement entropy \cite{Ryu:2006bv}, the 
first law for CFT entanglement entropy \cite{Blanco:2013joa,Wong:2013gua}, and how this follows from the bulk first law in the case of spheres \cite{Faulkner:2013ica}.  In section (\ref{extendedsection}) we begin by reviewing the construction of the extended first law for AdS black holes \cite{KastorEtal:2009} and then show how this can be applied to the minimal surface $\Sigma$ to extend the first law for holographic entanglement entropy.  Section (\ref{extendedsection}) also includes comments on how the variation of the bulk cosmological constant relates to varying the number of degrees of freedom in the boundary CFT and the natural interpretation of this in terms of chemical potential.  Section (\ref{conclude}) is devoted to concluding remarks regarding  future directions for exploration on this topic.

\section{Holographic entanglement entropy}\label{holosection}

Entanglement entropy is defined whenever the Hilbert space of a quantum theory can be split into two disjoint pieces.  In a quantum field theory, this can be done by splitting the spatial volume on a given constant time slice into two complementary regions $A$ and $B$.  The reduced density matrix associated with the region $A$ for a system described by the wavefunction $|\psi\rangle$ is obtained by tracing the 
full density matrix over the degrees of freedom in $B$
\begin{equation}
\rho_A = Tr_B |\psi\rangle\langle\psi |
\end{equation}
The entanglement entropy between the regions $A$ and $B$ is then given by the trace over ${\cal H}_A$
\begin{equation}
S_E=-Tr\rho_A\log\rho_A
\end{equation}
Ryu and Takyanagi have conjectured that the entanglement entropy in a CFT is given by equation (\ref{entropy}) where $\Sigma$ is a bulk minimal surface, whose boundary at spatial infinity matches on to the boundary between the regions $A$ and $B$.

\begin{SCfigure}
%\begin{center}
\includegraphics[width=0.4\textwidth]{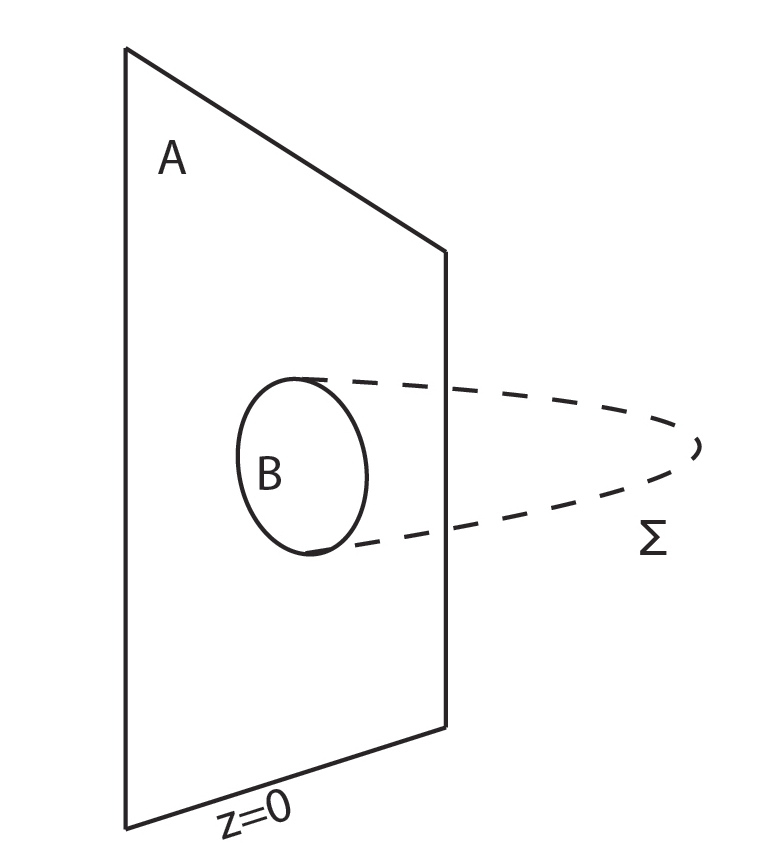}
%\end{center}
\caption{{\normalsize Constant time slice on the boundary $z=0$ divided into two disjoint regions $A$ and $B$ with corresponding minimal surface $\Sigma$ in the bulk AdS spacetime.}} 
\label{bulksurface}
\end{SCfigure}

In the case that $B$ is a spherical ball, the surface $\Sigma$ can be solved for exactly and the resulting holographic formula for the entanglement entropy computed.
Let us use the Poincare metric for $D$-dimensional AdS spacetime
\begin{equation}\label{metric}
ds_D^2= {l^2\over z^2}(dz^2-dt^2+d\vec x\cdot d\vec x)
\end{equation}
where spatial infinity is located at $z=0$ and the AdS radius $l$ is related to the cosmological constant $\Lambda$ by  
\begin{equation}\label{cosmo}
\Lambda = - {(D-1)(D-2)\over 2l^2}
\end{equation}
Take $B$ to be a spherical ball of radius $r_0$ on a constant time slice at infinity.  For notational simplicity we take $B$ to be centered at the origin and the constant time to be $t=0$.  It is straightforward to allow for an arbitrary spatial center for $B$ or constant time slice.  The corresponding bulk minimal surface $\Sigma$ on the $t=0$ hypersurface is then given by
\begin{equation}\label{sigma}
z^2 + r^2 = r_0^2
\end{equation}
where $r^2=\vec x\cdot\vec x$.  We see that the surface extends in the bulk to $z=r_0$.  The minimal surface $\Sigma$ is depicted in figure (\ref{bulksurface}).
The area of $\Sigma$ is given by
\begin{equation}\label{spherearea}
A_\Sigma = l^{D-2}\Omega_{D-3}\int_{y_c}^1dy {(1-y^2)^{D-4\over 2}\over y^{D-2}}
\end{equation}
where $y=z/r_0$ and a cutoff at $z_c$ has been imposed to regularize the area.  Note that $A_\Sigma$ depends on the AdS radius $l$ only through the overall prefactor of $l^{D-2}$.  The integral may be evaluated in any particular dimension, and gives for example
\begin{eqnarray}
A_\Sigma &=&2l\left (-\log y_c +\log(1+\sqrt{1-y_c^2})\right),\qquad D=3 \\
A_\Sigma &=&2\pi l^3\left\{ {\sqrt{1-y_c^2}\over y_c^2}+\log y_c-\log(1+\sqrt{1-y_c^2})\right\},\qquad D=5
\end{eqnarray}
displaying logarithmic and  quadratic divergences respectively as the cutoff $y_c$ goes to zero.

The first law for entanglement entropy \cite{Blanco:2013joa,Wong:2013gua} comes about in the following way.  The reduced density matrix $\rho_A$ may be used to define an effective thermal system via
\begin{equation}\label{thermal}
\rho_A = {e^{-H_A}\over Tr(e^{-H_A})}
\end{equation}
where $H_A$ is known as the modular Hamiltonian.  Under a change in the quantum state, the entanglement entropy then satisfies the first law type relation
\begin{equation}\label{firstent}
\delta S_E = \delta\langle H_A\rangle
\end{equation}
The explicit form of the modular Hamiltonian is not known for arbitrary shaped entangling surfaces.  In the special case of the sphere, however, it is given by
\begin{equation}\label{modular}
H_A = 2\pi\int_B d^{D-2}x\,  {r_0^2-r^2\over 2r_0}\, T_{tt}^{CFT}(t=0,\vec x)
\end{equation}
where $T_{\mu\nu}$ is the stress tensor of the boundary CFT and the integration is over the interior of the spherical ball $B$.  

\begin{SCfigure}
%\begin{center}
\includegraphics[width=0.4\textwidth]{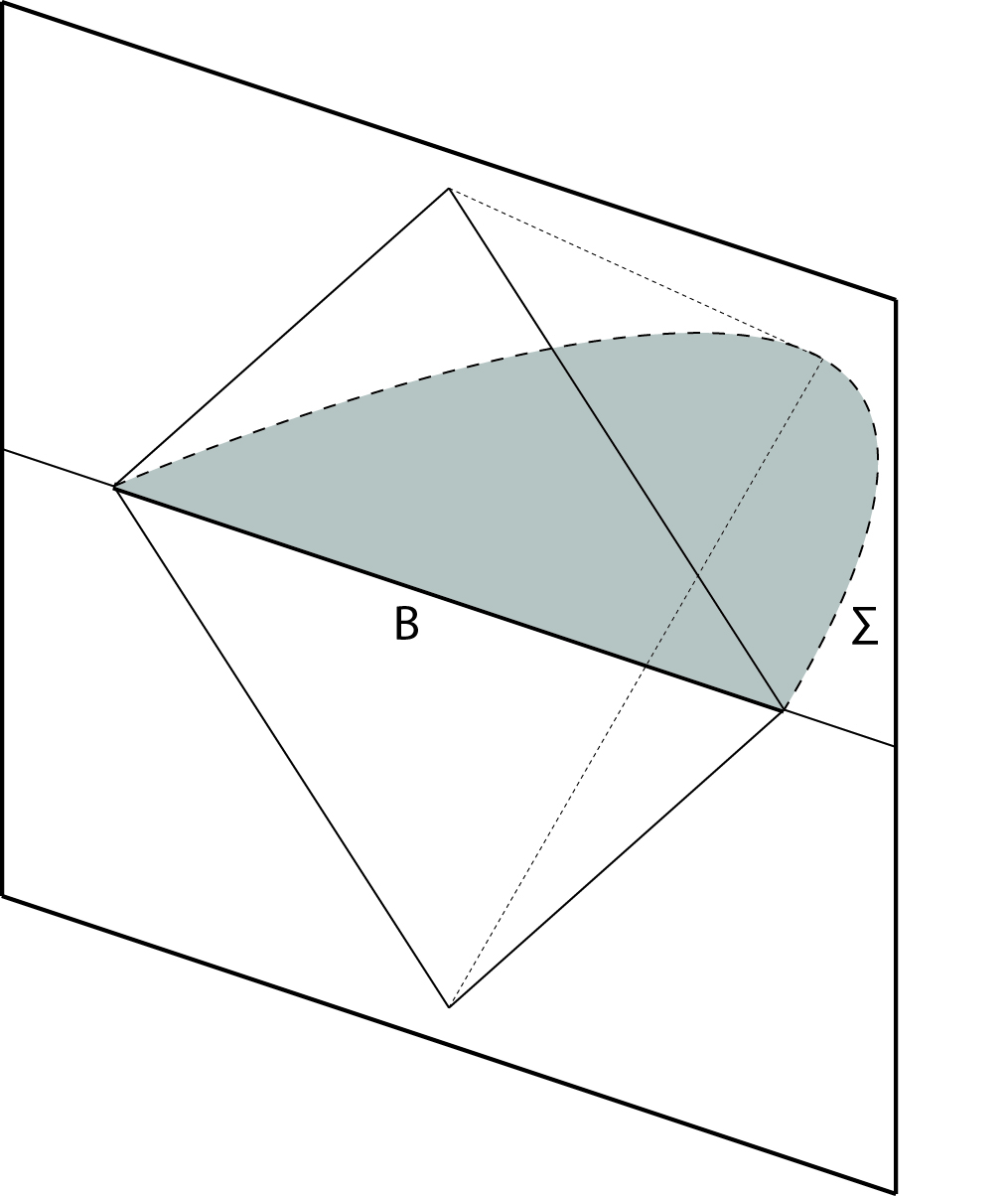}
%\end{center}
\caption{{\normalsize The causal developments of $B$ and $\Sigma$ are shown.  Time runs in the vertical direction and the spherical region $B$ appears as a line segment on the boundary at $z=0$.}} 
\label{causalfig}
\end{SCfigure}

In the case that $B$ is a spherical ball, it was also shown that the first law for entanglement entropy (\ref{firstent}) can be obtained from a bulk gravitational first law \cite{Faulkner:2013ica}.  As noted in the introduction, this is possible because the corresponding bulk minimal surface $\Sigma$ is the bifurcation surface of a Killing horizon, and the proof of the bulk first law given in \cite{SudarskyWald:1992} applies in this non-black hole context as well.  The bulk Killing vector in question is given by
\begin{equation}\label{killingvector}
\xi = -{2\pi\over r_0}(tz\partial_z +tx^k\partial_k)+{\pi\over r_0}(r^2_0-z^2-r^2-t^2)\partial_t
\end{equation}
and its norm is given by
\begin{equation}
\xi_a\xi^a = -{l^2\pi^2\over z^2r_0^2}[(r_0-t)^2-(r^2+z^2)][(r_0+t)^2-(r^2+z^2)]
\end{equation}
The Killing horizon is the null surface in the bulk where the norm vanishes. Its boundary at $z=0$ includes the causal diamond shown in figure (\ref{causalfig}).  The lower portion of the causal diamond is the set of all points on the boundary whose causal future on the boundary at time $t=0$ is included in $B$, while the upper portion is the set of points whose entire causal past on the boundary is in $B$.  The Killing vector $\xi$ is easily seen to vanish on the minimal surface $\Sigma$, given by (\ref{sigma}) on the $t=0$ hypersurface, and so it is the bifurcation surface for this Killing horizon.

The method of proving the bulk first law will be discussed in some detail below.  For the Killing vector $\xi$ given in (\ref{killingvector}) the result can be written as
\begin{equation}\label{firstlaw}
{\kappa \delta A_\Sigma\over 8\pi G}=\delta E_\xi
\end{equation}
where $\kappa$ is the surface gravity of the horizon, $\delta A_\Sigma$ is the change in area of the minimal surface $\Sigma$ under a perturbation to a nearby solution of the bulk equations of motion which keeps $\partial B$ fixed, 
 and $\delta E_\xi$ is the change in the ADM charge associated with the Killing vector $\xi$.    The surface gravity of $\xi$
  is found to be $\kappa=2\pi$.  Given the holographic identification of the entanglement entropy in (\ref{entropy}), the left hand side of (\ref{firstlaw}) is simply the change in the entanglement entropy of $B$ under this perturbation.  The ADM charge associated with $B$ was found in \cite{Faulkner:2013ica} to be 
\begin{equation}\label{killingcharge}
E_\xi= 2\pi\int_B d^{D-2}x {r_0^2-r^2\over 2r_0} T_{tt}^{boundary}(t=0,\vec x)
\end{equation}
where $T_{\mu\nu}^{boundary}$ is the boundary stress tensor.  According to the AdS-CFT correspondence we have $T_{\mu\nu}^{boundary}=\langle T_{\mu\nu}^{CFT}\rangle$ and therefore $Q_\xi$ is equal to the expectation value of the modular Hamiltonian (\ref{modular}).  The right hand side of the bulk first law (\ref{firstlaw}) then also maps to the right hand side of the first law for entanglement entropy (\ref{firstent}).

\section{Extended first law for entanglement entropy}\label{extendedsection}

In this section we establish an extension of the first law of entanglement entropy (\ref{firstent})  that includes variations in the number of degrees of freedom of the boundary theory.  
We follow the approach of \cite{Faulkner:2013ica} outlined above, starting with an extension of the bulk gravitational first law that includes variation in the cosmological constant \cite{KastorEtal:2009} and applying this in the context of the holographic  formula for entanglement entropy \cite{Ryu:2006bv}.  Our result is  thus limited to the case of spherical entangling surfaces on the boundary and by the validity of the holographic prescription for computing entanglement entropy.  We will comment on ideas for extending this approach to arbitrary shaped entangling surfaces in the conclusion.

The extension of the bulk first law \cite{KastorEtal:2009} was itself established using  the Hamiltonian perturbation theory techniques that were used in the proof 
\cite{SudarskyWald:1992} of the ordinary first law for black hole thermodynamics\footnote{See \cite{Traschen:1985} for an application of these techniques in a cosmological context.}.  
In order to see how the extended bulk first law applies in the context of holographic entanglement entropy, we present a brief sketch of this construction.  The basic idea is that for perturbations around a solution to the gravitational field equations with a Killing symmetry, {\it e.g.} a stationary black hole solution, the perturbed Hamiltonian constraint equations can be combined into a Gauss' law type statement.  Integrating over the region between the black hole horizon and spatial infinity then establishes an equality between a boundary integral at infinity, given in terms of variations in the ADM charges corresponding to the Killing symmetry, and an integral at the horizon, which is given in terms of the variation in the horizon area.  In the holographic entanglement entropy context, the inner boundary becomes the minimal surface $\Sigma$, while the outer boundary is over the region $B$ at spatial infinity.

\subsection{Bulk first law via Hamiltonian perturbation theory}
 We start by assuming that we have a foliation of a spacetime by a family of hypersurfaces 
denoted by $S$.
The unit timelike normal to the hypersurfaces is written as $n^a$ with $n\cdot n=-1$. The spacetime metric is given by
\begin{equation}\label{metricsplit}
g_{ab} =-n_a n_b + s_{ab}\,,
\end{equation}
where the metric $s_{ab}$ on the hypersurfaces $S$ satisfies $s_a{}^b n_b =0$.
The dynamical variables in the Hamiltonian formalism  for general relativity are
the  metric $s_{ab}$ and its canonically conjugate momentum $\pi ^{ab} =-\sqrt{s} (K^{ab}-Ks ^{ab} )$,  where  $K_{ab} =s_a{}^c \nabla _c n_b$
is the extrinsic curvature of a hypersurface $S$ and $K$ is the trace $K=K^a{}_a$. 
We consider Hamiltonian evolution
along a vector field $\xi ^a$, which can be decomposed into its components normal and tangential to $ S$ according to
\begin{equation}\label{xidecomp}
\xi^a=Fn^a +\beta^a\,,
\end{equation}
with $F=-\xi\cdot n$ giving the lapse function and $\beta^a$ the shift vector.
The gravitational Hamiltonian 
for evolution of
the system along $\xi^a$ is then given by $H_{grav}  =FH+\beta^a H_a$ where
\begin{eqnarray}
H&\equiv& -2G_{ab} n^a n^b =  -\, R^{(D-1)}  + {1 \over |s|}  \Bigl({\pi ^2 \over D-2 } - \pi^{ab} \pi_{ab} \Bigr)\,,\label{hamconstraint}\nonumber\\
H_b&\equiv& -2G_{ac} n^a s^c _b=-2\,  D_a (|s|^{-{1 \over 2}} \pi^{ab} )\,.\label{momconstraint}
\end{eqnarray}
are the Hamiltonian and momentum constraint operators.  
Here $R^{(D-1)}$ is the scalar curvature for the metric $s_{ab}$ and $D_a$ is the derivative operator on the hypersurface $S$.
If the spacetime satisfies the Einstein equations with a cosmological constant  $G_{ab} =-\Lambda g_{ab} $ then the Hamiltonian and momentum  constraint
equations imply that 
\begin{equation}\label{constraint}
H =- {2} \Lambda\,,  \quad H_b =0\,.
\end{equation}
Let us now assume that our spacetime with metric  $g_{ab}$ solves the Einstein equation with cosmological constant $\Lambda$ and has a symmetry generated by the Killing vector
$\xi^a$.
We further assume that the metric $\tilde g_{ab} =g_{ab} +\delta g_{ab}$ is the linear approximation to 
another nearby solution to the  Einstein equations with a perturbed cosmological constant 
$\Lambda +\delta\Lambda$. 
The Hamiltonian data for this perturbed metric are $\tilde s_{ab}=s_{ab}+\delta s_{ab}$ and $\tilde\pi^{ab}=\pi^{ab} +\delta\pi^{ab}$, where the $s_{ab}$, $\pi^{ab}$ are the unperturbed data.  Below we will use the notation $h_{ab}=\delta s_{ab}$ and $p^{ab}=\delta \pi ^{ab} $ for the perturbations to the spatial metric and momentum.   The corresponding linearized Hamiltonian and momentum constraint equations are given by $H+\delta H = -2(\Lambda +\delta\Lambda)$ and $H_b+\delta H_b=0$.
Together with the unperturbed constraint equations (\ref{constraint}) these imply that the perturbations to the Hamiltonian and momentum constraint operators satisfy
\begin{equation}\label{perturb}
\delta H= -2\delta\Lambda,\qquad \delta H_b=0
\end{equation}

The key step is now to note \cite{Traschen:1985,SudarskyWald:1992,TraschenFox:2004} that
contracting the linearized constraints $(\delta H,\delta H_a)$ with the normal and tangential components of the Killing vector $(F,\beta^a)$ gives a total derivative on the hypersurface,
\begin{equation}
F\delta H + \beta^a\delta H_a= D_a B^a.
\end{equation}
If the extrinsic curvature of the slice vanishes, which will be the case of most interest to us, then the boundary vector $B^a$ is given by\footnote{In the general case of non-vanishing extrinsic curvature, the boundary vector $B^a$ also includes the additional term
 $${1 \over \sqrt{|s|} } \beta^b (\pi^{cd} h_{cd} s^a{}_b - 2 \pi^{ac} h_{bc} -2p^a{}_b ).$$ }
\begin{equation}\label{gaussvector}
 B^a  =   F(D^a h - D_b h^{ab})  - h D^a F + h^{ab} D_b F 
\end{equation}
Since the perturbations to the spatial metric and momentum $h_{ab}$,  $p^{ab}$ solve the linearized  constraint equations (\ref{perturb}), the vector $B^a$ 
therefore satisfies the Poisson type equation
\begin{equation}\label{gauss}
D_a B^a =  - {2} F\delta \Lambda\,.
\end{equation}  
 The fact that  $\xi^a$ is a Killing vector implies that the source term on the right hand side
 may also be written as a total derivative and moved to the left hand side \cite{Kastor:2008xb,KastorEtal:2009}.  One defines a Killing potential $\omega^{ab}$
 associated with a Killing vector $\xi ^a$ to be an antisymmetric tensor that satisfies the equation
 \begin{equation}\label{omegadef}
 \nabla _c \omega ^{cb} = \xi ^b\,.
 \end{equation}
 The existence of a Killing potential is guaranteed since $\nabla_a\xi^a=0$, while  non-uniqueness of the Killing potential cancels out from the construction.
 The normal component of $\xi$
 can then be written in terms of the Killing potential as
$F=-D_c (\omega ^{cb} n_b )$
and we obtain the Gauss' law type statement  
 \begin{equation}\label{altgauss}
 D_a ( B^a   -{2}\delta \Lambda \omega ^{ab} n_b ) =0\,.
 \end{equation}
 
First, let us set $\delta\Lambda=0$ and see how the ordinary first law comes about in a static black hole spacetime.
   In that case equation (\ref{altgauss}) reduces to $D_aB^a=0$ and we integrate this statement over a spatial hypersurface that stretches from the black hole horizon $\cal H$ in the interior out to spatial infinity.  Using Gauss' theorem we then have the relation
$\int_\infty da_a B^a-\int_{\cal H}da_a B^a=0$. Choosing $\xi$ to be the static Killing field, 
the integral at the horizon is proportional to the change in horizon area under the perturbation
\begin{equation}\label{deltaa}
\int_{\cal H}da_aB^a=-2\kappa\delta A_{BH}
\end{equation}
 and the boundary term at infinity is proportional to the change in the ADM mass
\begin{equation}\label{infinity}
\int_\infty da_aB^a = -16\pi G\delta M
\end{equation}
Combining these results yields the ordinary first law for black hole horizons $\delta M= \kappa\delta A_{BH}/8\pi G$.

The derivation of the first law for CFT entanglement entropy for spherical regions on the boundary 
from the bulk first law in \cite{Faulkner:2013ica} makes use of this same formalism.  In this case one integrates $D_aB^a=0$ over the shaded region of the constant time $t=0$ slice in the bulk shown in figure (2).  This is bounded by the minimal surface $\Sigma$ in the interior and at spatial infinity by the spherical ball $B$.  The boundary vector $B^a$ is that formed from the Killing vector $\xi$ in (\ref{killingvector}).  In this case, the resulting boundary integral at $\Sigma$ is given, similarly to (\ref{deltaa}), by $-2\kappa\delta A_\Sigma$.  For the boundary integral at infinity, one finds that  the integrand is equal to the integrand for the ADM mass $\delta M$ but weighting by an additional factor of $\pi  (r_0 ^2 - r^2  )/r_0$.  The integrand for $\delta M$ coincides with the time-time component of the boundary stress tensor $T^{boundary}_{tt}$.  The boundary integral at infinity is then given by a factor of minus $16\pi$ times the variation of the ADM charge $E_\xi$ in (\ref{killingcharge}) yielding the first law type relation (\ref{firstlaw}).

Let us now consider the case $\delta\Lambda\neq 0$, where we perturb the cosmological constant \cite{KastorEtal:2009}.  For definiteness let us assume that $\Lambda<0$.   Integrating the divergence relation (\ref{altgauss}) over the region between the black hole horizon and spatial infinity, then applying Gauss' theorem now leads to the statement
\begin{equation}
\int_\infty da_a (B^a- 2\omega^{ab}n_b\delta\Lambda)-\int_{\cal H}da_a (B^a- 2\omega^{ab}n_b\delta\Lambda)=0
\end{equation}
One can then evaluate the different parts of this relation.  The integral of the boundary vector $B^a$ over the horizon is still given by (\ref{deltaa}).  The integral of the boundary vector $B^a$ at infinity, however, is more complicated.  The boundary vector (\ref{gaussvector}) depends on the metric perturbation $h_{ab}$. 
To derive the usual first law for black holes, one must take
 standard asymptotically AdS fall off  conditions for $h_{ab}$ (see e.g. \cite{Kastor:2008xb,KastorEtal:2009}),
 and the integral (\ref{infinity}) then yields a finite change in the ADM mass.  With $\delta\Lambda\neq 0$, however, the asymptotics of the perturbed metric are shifted relative to the background, causing the integral of $B^a$ over the boundary at spatial infinity to include a divergence proportional to $\delta\Lambda$.   
 One finds that the change in the ADM mass is given by the subtracted integral
\begin{equation}\label{deltam}
\int_\infty da_a(B^a-2\, \omega^{ab}_{AdS}\, n_b\delta\Lambda)=-16\pi G\delta M
\end{equation}
where $\omega^{ab}_{AdS}$ is the Killing potential in the background AdS spacetime with cosmological constant $\Lambda$.  The resulting extended first law then has the form
\begin{equation}\label{bhfirst}
\delta M = {\kappa\delta A_{BH}\over 8\pi G} +{\Theta\delta\Lambda\over 8\pi G}
\end{equation}
where the thermodynamic potential associated with varying the cosmological constant is given by
\begin{equation}\label{thermovolume}
\Theta = \int_{\cal H} da_a\omega^{ab}n_b - \int_\infty da_a(\omega^{ab}-\omega^{ab}_{AdS})n_b \, .
\end{equation}
Adding and subtracting  the background Killing potential to the integrand serves to make both $\Theta$ and $\delta M$ finite.
 
 \subsection{Extended first law for entanglement entropy}
 
The formalism developed in the last section is applicable in the holographic entanglement entropy context as well as for black hole horizons.
  In this case, following  \cite{Faulkner:2013ica},
we take the region of integration $U$ to be the volume bounded by the bulk minimal surface $\Sigma$ in the interior, out to the portion of spatial infinity covered by the spherical ball $B$.
Integrating over the region $U$ then yields the relation
\begin{equation}\label{gaussint}
\int_{\Sigma} da_c \left( B^c   - {2}\delta \Lambda \omega ^{cb} n_b \right) 
 - \int_{B} da_c \left( B^c   - {2} \delta \Lambda \omega ^{cb} n_b \right)=0\,.
\end{equation}
the area element $da_c$ is taken to  point into the region $U$ on the minimal surface $\Sigma$ in the interior and out of
$U$ on the ball $B$ at spatial infinity.  The boundary vector and Killing potential depend on the Killing vector $\xi^a$ given in (\ref{killingvector}),
which vanishes on $\Sigma$.
The Killing potential may be easily found by combining the identity for Killing vectors $\nabla_a\nabla^a\xi^b=-R^b{}_c\xi^c$ with Einstein's equation 
$G_{ab}=-\Lambda g_{ab}$ for the AdS background to get\footnote{The Killing potential is not uniquely defined.   However, as discussed in \cite{KastorEtal:2009} this non-uniquenss does not affect the construction. For example, a term of the form $\nabla_c\lambda^{abc}$ where $\lambda^{abc}$ is a totally antisymmetric tensor can always be added to a Killing potential to get another Killing potential.  However, such a term will integrate to zero in (\ref{gaussint}).}
\begin{equation}
\omega^{ab} = -{D-2\over 2\Lambda}\nabla^a\xi^b
\end{equation}
For the Killing vector (\ref{killingvector}) this gives
\begin{equation}\label{killingpot}
\omega = {1\over 2}\omega^{ab}\partial_a\wedge\partial_b = {\pi z\over (D-1)r_0}\left\{ (r_0^2+z^2-t^2-r^2)\partial_t\wedge\partial_z +
2tx^k\partial_z\wedge\partial_k + 2zx^k\partial_t\wedge\partial_k\right\}
\end{equation}

The task is now to evaluate the different boundary integrals appearing in (\ref{gaussint}).    At infinity, as in the case with the black hole spacetime, it is necessary
to appropriately group terms in the integral to have finite, geometrical, combinations. Analogous to equation (\ref{deltam}) for $\delta M$,
which is associated with the static Killing field,  the combination
that gives a finite ADM charge $E_\xi$ associated with the Killing vector $\xi^a$ is
\begin{equation}\label{inftyintegral}
\int_\infty da_a(B^a-2\, \omega^{ab}_{AdS}\, n_b\delta\Lambda)=-16\pi G\delta E_\xi
\end{equation}
One finds that the Killing potential term exactly cancels the new term in the boundary vector $B^a$ that arises from varying the cosmological constant.  
This cancelations comes about as follows.
From (\ref{cosmo}) a variation in $\Lambda$ is related to a variation in the AdS radius by
\begin{equation}
\delta\Lambda = {(D-1)(D-2)\over l^3}\delta l
\end{equation}
We will work here in terms of $\delta l$.
Since the first law construction is linear in perturbations, in order to construct the new terms in the first law when $l$ is varied, we can consider this variation in isolation.
The variation to the AdS metric (\ref{metric}) is then simply $\delta g_{ab} = (2\delta l/l)g_{ab}$ giving $h_{ab}=(2\delta l/l)s_{ab}$ for the perturbation to the spatial metric.  The normal component $F$ of the Killing vector $\xi^a$ in the expression (\ref{gaussvector}) for the boundary vector $B^a$ is given by $F= (\pi l/r_0 z)(r_0^2-z^2-r^2)$.  The area element at the boundary at spatial infinity points in the $z$-direction and hence one needs only the $z$-component of $B^a$ which is found to be
\begin{equation}
B^z = {2(D-2)\pi\delta l\over r_0 l^2}(r_0^2+z^2-r^2)
\end{equation}
On the other hand, one finds that the $z$ component of the Killing potential term in (\ref{inftyintegral}) is also given by
\begin{equation}\label{deltal}
2\omega^{zt}n_t\delta\Lambda = {2(D-2)\pi\delta l\over r_0 l^2}(r_0^2+z^2-r^2)
\end{equation}
leading to the cancelation.    Here we have used the fact that our background spacetime is unperturbed AdS, so that $\omega^{ab}_{AdS}=\omega^{ab}$ and is given by (\ref{killingpot}).
Hence the the boundary integral (\ref{inftyintegral}) at infinity receives no new contributions from varying the cosmological constant
The integral of the boundary vector $B^a$ over the minimal surface $\Sigma$ is again given by 
\begin{equation}\label{deltaA}
\int_{\Sigma}da_aB^a=-2\kappa\delta A_{\Sigma}
\end{equation}
with surface gravity $\kappa=2\pi$ for the Killing vector $\xi^a$.

Combining theses results, we then have the extended bulk first law 
\begin{equation}\label{extended}
\delta E_\xi = {\delta A_\Sigma\over 4G} -{V\delta\Lambda\over 8\pi G}
\end{equation}
where the thermodynamic volume is given by\footnote{The boundary term at spatial infinity cancels in (\ref{thermovolume}) because, as noted above, we are perturbing about AdS spacetime and therefore $\omega^{ab}_{AdS}=\omega^{ab}$.} 
\begin{equation}\label{vol}
V\equiv -\Theta = -\int_\Sigma da_a\omega^{ab}n_b
\end{equation}
This integral may be evaluated using the explicit forms for the unit normal to the constant time slice $n= -(l/z)dt$ and $da_b=m_b\, da $ where
\begin{equation}
m = -{l\over zr_0}(zdz +\vec x\cdot d\vec x)
\end{equation}
is the outgoing normal to $\Sigma$ within the constant time slice and $da$ is the induced area element.
One then finds that\footnote{The thermodynamic volume in this case has dimension of $(length)^D$ rather than $(length)^{D-1}$ as a true volume would.  This stems from the normalization of the Killing vector $\xi$ in (\ref{killingvector}), which makes it dimensionless, rather than having the usual scaling of $(length)^{-1}$ as {\it e.g.} for the time translation Killing vector $\partial_t$.}
\begin{equation}\label{Theta}
V = {2\pi l^2\over D-1}\, A_\Sigma
\end{equation}
where $A_\Sigma$ is the area of the minimal surface $\Sigma$ given explicitly in (\ref{spherearea}).

One can check that this new contribution to the first law is correct by looking at the explicit form for the area $A_\Sigma$ given in (\ref{spherearea}).
We see that the dependence on the AdS radius comes only through the overall prefactor of $l^{D-2}$.  If we consider the extended first law (\ref{extended}) in the case that only  the cosmological constant is varied, so that $\delta E_{\xi}=0$, then we find 
\begin{equation}
\delta A_\Sigma = (D-2)A_\Sigma{\delta l\over l}
\end{equation}
Making use of (\ref{deltal}) and (\ref{Theta}) we then see that the extended first law (\ref{extended}) is satisfied.  

Finally, we can rewrite the extended first law (\ref{extended}) with the thermodynamic volume (\ref{vol}) entirely in terms of the entanglement entropy $S_E$ and the AdS curvature radius $l$ as
\begin{equation}\label{extended2}
\delta E_\xi = \delta S_E - (D-2)S_E{\delta l\over l}
\end{equation}
We see that the simple scaling of the holographic entanglement entropy with $l$ displayed by the exact formula (\ref{spherearea}) can be interpreted as following from the extended first law with the thermodynamic potential $\Theta$ determined geometrically by the integral (\ref{vol}) of the Killing potential over the bulk minimal surface $\Sigma$.

 \subsection{CFT interpretation}

In the last section we found an extension of the bulk first law (\ref{extended}) for the minimal surface $\Sigma$ 
that includes varying the cosmological constant $\Lambda$, or equivalently
the AdS curvature radius $l$.  Following the logic of \cite{Faulkner:2013ica} this
implies a similar extension (\ref{extended2}) of the first 
law for the CFT entanglement entropy for spherical regions on the boundary.  
In this section we discuss the significance of this result in the boundary CFT. The extended first law
provides an interpretation of the thermodynamic
volume, which was originally introduced as a gravitational quantity \cite{KastorEtal:2009}, in terms of the CFT quantities.  We see from (\ref{extended}) that $V$ is proportional to
the the variation of the entanglement entropy with respect to $l$ with $E_\xi$ fixed.
Generally speaking, the AdS curvature radius in the bulk is a measure of the number of degrees of freedom, $\nbar$, of the boundary field theory, with large AdS radius $l$ corresponding to the limit of large $\nbar$.  Varying $l$ in the bulk then corresponds to varying $\nbar$ in the boundary theory, {\it i.e.} to varying the boundary CFT itself within a family of such theories having different numbers of degrees of freedom, so that 
\begin{equation}\label{volmeans}
V= {8\pi l^3 \over (D-2)(D-1) } {\delta S_E \over \delta l} \propto {\delta S_E \over \delta \nbar}
\end{equation}
 The precise correspondence between $l$ and $\nbar$, however, depends on the family of CFT's  under consideration.
      
For even dimensional boundary CFT's, corresponding to odd bulk dimensions $D$, the correspondence between the bulk AdS radius of curvature and properties of the boundary theory can be framed in terms of conformal anomalies \cite{Henningson:1998gx}.
For example, with a $D=3$ dimensional bulk, the boundary theory is a $2$-dimensional CFT and the conformal anomaly is proportional to the Virasoro central charge $c$.  The AdS curvature radius $l$ and the central charge are related according to \cite{Brown:1986nw}
\begin{equation}
c={3l\over 2G}
\end{equation}
The extended first law (\ref{extended2}) then translates into the relation for the boundary theory
\begin{equation}\label{extended3}
\delta E_\xi = \delta S_E - {S_E\over c}\delta c.
\end{equation}
Note that because of the logarithmic derivative, factors of Newton's constant disappear, so that the relation only depends on quantities in the boundary theory.  The central charge is normalized so that $c=1$ for a massless scalar field on the boundary, while a theory of $N$ noninteracting, massless scalars has $c=N$.  The extended first law then gives the dependence of the entanglement entropy on the number of degrees of freedom $N$.  

Recall that the chemical potential $\mu$ of a thermodynamic system is the thermodynamic variable conjugate to a change in particle number $N_{part}$ and appears in the first law as 
\begin{equation}\label{chemfirst}
dE = TdS +\mu\, dN_{part}.
\end{equation}
It is natural to think of the new term in the first law (\ref{extended3}) as a chemical potential term for varying the number of degrees of freedom as measured by the central charge $c$, with the chemical potential given by 
\begin{equation}
\mu_{CFT} = -{S_E\over c}.
\end{equation}
For comparison, the chemical potential for an ideal gas is given by 
\begin{equation}
\mu_{ideal} = -T\left({S\over N_{part}}-{5\over 2}\right).
\end{equation}  
Recognizing that the effective thermodynamic ensemble for the entanglement entropy (\ref{thermal}) is defined to have unit temperature, we see that the chemical potential $\mu_{CFT}$ for varying $c$ in the context of entanglement entropy matches the first term in the chemical potential of an ideal gas,
that is, minus the entropy over the number of degrees of freedom.   This term reflects an overall extensivity of the entropy on particle number, while
the minus sign arises because at fixed energy, the entropy increases
if $N$ is increased.

A similar story holds in higher, even number of boundary dimensions.  For example, a holographic calculation of the conformal anomaly for a
$4$-dimensional boundary theory with an Einstein gravity dual gives in the notation of \cite{Henningson:1998gx}
\begin{equation}
{\cal A} = {l^3\over 8\pi G_5}\left({1\over 8}R_{\mu\nu}R^{\mu\nu}-{1\over 24}R^2\right)
\end{equation}
where $R_{\mu\nu\rho}{}^\sigma$ is the curvature of the boundary metric.  This can be matched to a field theory calculation of the conformal anomaly in order to find the relationship between the AdS curvature radius and the number of boundary degrees of freedom.  Note, however, that the extended first law (\ref{extended2}) holds in all dimensions, both even and odd, and hence such a  relation between the AdS radius $l$ and the conformal anomaly of the boundary theory does not cover all cases of interest.

The most familiar example with a $4$ dimensional boundary theory is ${\cal N}=4$ $SU(N)$ SYM theory.  The conformal anomaly in this theory can be computed approximately using free fields, yielding in the large $N$ limit the identification $l^3 = 2 G_5 N^2/ \pi$.  
However, in the AdS/CFT correspondence arising from compactifying $10$-dimensional Type IIB string theory on $AdS_5\times S^5$ with $N$ units of $5$-form flux \cite{Maldacena:1997re}, the $5$-dimensional Newton's constant depends on the AdS radius as well through $G_5=G_{10}/\pi^3 l^5$, giving the net dependence of $l$ on $N$
\begin{equation}
l^8 =  {2G_{10} N^2\over \pi^4}
\end{equation}
where $G_{10}$ is the fixed $10$-dimensional Newton's constant,  so that $\delta l/l=\delta N/4N$.    To apply the extended first law in this case, we need to take into account the explicit dependence of the entanglement entropy on $G_5$ in equation (\ref{entropy}) as well, which gives
\begin{equation}
\delta S_E= {1\over 4G_5}\delta A_\Sigma -{A_\Sigma }{\delta G_5 \over G_5 }
\end{equation}
Combining this with the extended first law (\ref{extended2}) gives the net result for  ${\cal N}=4$ $SU(N)$ SYM theory
\begin{equation}\label{first4}
\delta E_\xi =\delta S_E-2S_E{\delta N\over N}
\end{equation}
In comparison with (\ref{chemfirst}), we can think of the coefficient of the new term in (\ref{first4}) as a chemical potential for varying the rank $N$ of the gauge group given by
\begin{equation}
\mu_N = - {2S_E\over N}
\end{equation}
where the additional factor of $2$ relative to $\mu_{ideal}$ arises because the number of degrees of freedom grows like $N^2$.
Similar analyses can be made for $AdS_7\times S^4$ and $AdS_4\times S^7$ which arise from the near horizon limits of stacks of $N$ $M2$ or $M5$-branes \cite{Maldacena:1997re} and provide similar formulas for the AdS radius and lower dimensional Newton's constant in terms of the number of branes $N$.

\section{Conclusions}\label{conclude}

We have shown how the extension of the first law to include variation in the cosmological constant \cite{KastorEtal:2009} can be applied in the AdS/CFT context to yield an extension of the first law for CFT entanglement entropy.  This extension takes the form of a chemical potential term for varying the number of degrees of freedom on the boundary, with the chemical potential having a similar form to the thermodynamic chemical potential of an ideal gas.

Like the derivation of the ordinary un-extended first law for entanglement entropy from the bulk first law \cite{Faulkner:2013ica}, our result holds only for spherical entangling surfaces on the boundary.  The first law for CFT entanglement, on the other hand, has been shown to hold for general entangling surfaces \cite{Blanco:2013joa,Wong:2013gua}, although the explicit form of the modular Hamiltonian in (\ref{firstent}) is not generally known.  It is natural to ask whether there is some bulk construction that yields the boundary first law for a general entangling surface.  In the general case the bulk minimal surface 
$\Sigma$ in the holographic prescription for entanglement entropy will not be part of  a Killing horizon.  There has been some progress made on establishing an alternative formalism for the laws of black hole thermodynamics based on the properties of trapping horizons \cite{Hayward:1993wb}.  It would be interesting to see whether this formalism may be applied to the minimal surfaces that arise in the context of the holographic prescription for entanglement entropy.

A second, more straightforward, area for future investigation is to apply the extended first law formalism of this paper to bulk AdS black holes and the corresponding thermal states of CFT's.  Our construction applies to perturbations from AdS, and therefore to the far field regions of AdS black holes in the bulk.  If we limit our attention to small spherical surfaces on the boundary, so that the corresponding bulk minimal surfaces $\Sigma$ are entirely within the far field regime, then our methods will yield the dependence of the entanglement entropy for thermal boundary states on the number of boundary degrees of freedom.

Finally, more general boundary CFT's have bulk duals that include higher curvature interactions.  Already for $4$-dimensional boundaries, the Gauss-Bonnet interaction must be added to the bulk gravitational theory in order to reproduce a general conformal anomaly for the boundary CFT \cite{Nojiri:1999mh}.  The construction of \cite{Faulkner:2013ica} yielding the boundary first law for entanglement entropy from the bulk first law applied to Lovelock theories in the bulk as well.  Extension of the bulk first law to include variations of the full set of couplings of Lovelock theories were considered in \cite{Kastor:2010gq}.  As in the present paper, this should lead to similar extensions of the boundary first law for entanglement entropy for spherical surfaces.  The various Lovelock couplings distinguish between different aspects of the boundary CFT's.  Computing the corresponding thermodynamic potentials for variation of the entanglement entropy with respect to these couplings should yield insight into the corresponding properties of the boundary theories.

\subsection*{Acknowledgements}  The authors thank Alex Maloney for helpful conversations.  The work of S.R. is supported by FONDECYT grant 11110176 and CONICYT
grant 791100027.

\end{document}